\documentclass[twocolumn]{aastex7}
\usepackage{amsmath}
\usepackage{amssymb}
\usepackage{xspace}
\usepackage{enumitem}
\usepackage[dvipsnames]{xcolor}

\newcommand{\tess}{\emph{TESS}\xspace}
\newcommand{\kepler}{\emph{Kepler}\xspace}

\shorttitle{Warm Jupiter Dynamics}
\shortauthors{Dong, Lee, Kokubo, Murray-Clay, \& Gupta}

\received{March 20, 2026}
\submitjournal{AAS Journals}

\graphicspath{{./}{figures/}}

\begin{document}

\title{Planet--Planet Scattering Explains the Mass--Eccentricity Relation of Warm Jupiters}

\author[orcid=0000-0002-3610-6953,gname=Jiayin,sname=Dong]{Jiayin Dong}
\affiliation{Department of Astronomy, University of Illinois at Urbana-Champaign, Urbana, IL 61801, USA}
\affiliation{Center for Astrophysical Surveys, National Center for Supercomputing Applications, Urbana, IL 61801, USA}
\email[show]{jdongx@illinois.edu}

\author[orcid=0000-0002-1228-9820,gname=Eve,sname=Lee]{Eve J.~Lee}
\affiliation{Department of Astronomy \& Astrophysics, University of California, San Diego, La Jolla, CA 92093-0424, USA}
\email{evelee@ucsd.edu}

\author[0000-0002-5486-7828,gname=Eiichiro,sname=Kokubo]{Eiichiro Kokubo}
\affiliation{Center for Computational Astrophysics, National Astronomical Observatory of Japan, Osawa, Mitaka, Tokyo 181-8588, Japan}
\email{kokubo.eiichiro@nao.ac.jp}

\author[orcid=0000-0001-5061-0462,gname=Ruth,sname=Murray-Clay]{Ruth Murray-Clay}
\affiliation{Department of Astronomy and Astrophysics, University of California, Santa Cruz, 1156 High Street, Santa Cruz, CA 95064, USA}
\email{rmc@ucsc.edu}

\author[orcid=0000-0002-5463-9980,gname=Arvind,sname=Gupta]{Arvind Gupta}
\affiliation{U.S. National Science Foundation National Optical-Infrared Astronomy Research Laboratory, 950 N. Cherry Avenue, Tucson, AZ 85719, USA}
\email{arvind.gupta@noirlab.edu}

\begin{abstract}
Warm giant planets with orbital periods of tens of days exhibit a positive correlation between mass and eccentricity.
We interpret this trend as the outcome of planet--planet scattering, representing a transition from collision-dominated interactions among low-mass planets to ejection-dominated interactions among high-mass planets. 
This framework has important implications for warm Jupiter origins. It suggests that warm Jupiters originate from compact, multi-planet configurations. The dynamical interactions that shape their present-day architectures likely occur near their current semimajor axes, regardless of whether warm Jupiters formed through convergent disk-driven migration or in-situ formation.
We argue that several observed properties of warm Jupiter systems, including the eccentricity bimodality, the mass--eccentricity relation, and generally low stellar obliquities, can be explained by this picture.
We further predict that not only circular warm Jupiters, but also eccentric warm Jupiters, should frequently have additional planetary companions that are detectable through radial velocity observations.
Finally, scattering can produce eccentricities high enough to trigger high-eccentricity tidal migration, potentially explaining the emerging population of proto-hot Jupiters on tidal migration tracks.
\end{abstract}


\section{Introduction}

Long-period giant planets discovered via radial velocity (RV) surveys exhibit a positive correlation between planetary mass and orbital eccentricity \citep{Butler06, Wright09}.
These radial-velocity planets have orbital periods ranging from tens of days to thousands of days and projected masses ($M_p\sin{i}$; where $M_p$ is the true planet mass and $i$ is the orbital inclination) of roughly 0.1--10 $M_{\rm Jup}$.
The observed mass--eccentricity relationship has been widely interpreted as evidence of dynamical interactions among multiple planets in the same system \citep[e.g.,][]{Ford08, Chatterjee08, Juric08, Frelikh19}.

\cite{Frelikh19} proposed a holistic explanation for the mass and eccentricity distributions observed in radial velocity planet surveys.
They showed that the upper envelope of the eccentricity distribution is consistent with the theoretical limit from planet--planet scattering, defined by the ratio of a planet's surface escape velocity to the local Keplerian velocity.
Since the escape velocity scales with the square root of planet mass, more massive planets are expected to have a higher eccentricity envelope, naturally explaining the observed mass--eccentricity correlation.
In addition, the highest-eccentricity planets ($e > 0.6$) preferentially orbit metal-rich host stars ([Fe/H] $> 0$) \citep[e.g.,][]{Dawson13, Alqasim25}.
This correlation may suggest that super-Jupiter mass growth occurs through collisions.
With stellar metallicity serving as a proxy for total solid mass in the disk, metal-rich systems initially contain more raw material to support the formation of multiple giant planets and subsequent collisional evolution and eccentricity excitation.

Whether this mass--eccentricity relation extends to the shortest-period giant planets cannot be easily evaluated.
Most detections of short-period giant planets are hot Jupiters with orbital periods of just a few days.
Their eccentricities have been efficiently damped by tidal interactions with their host stars, thereby erasing any dynamical imprints \citep[see][for a review and references therein]{Dawson18}.
In contrast, warm Jupiters--giant planets with orbital periods of roughly 10 to 365 days--occupy a regime where tidal circularization is less effective, allowing their eccentricities to preserve clues about their dynamical histories. While the eccentricity-envelope analysis of \citet{Frelikh19} (see their Figure 2) relies on the dynamical behavior of warm Jupiters, they were unable to harness the full dynamical potential of this region due to a relatively small statistical sample and reliance on radial velocity masses that suffer from the $M_p\sin{i}$ degeneracy.  
The longer periods and lower transit probabilities of warm Jupiters (as compared to hot Jupiters) have historically made them more difficult to detect via transit surveys.

The advent of the Transiting Exoplanet Survey Satellite \citep[\tess;][]{Ricker15} has significantly expanded the sample of known warm Jupiters in recent years, aided by observing strategies such as continuous viewing zones and repeated visits to the same regions of the sky. 
\citet{2024Natur.632...50G} first identified a positive correlation between true planetary mass and eccentricity in a sample of 52 transiting warm Jupiters.
Owing to their transiting nature, these systems do not suffer from the $M_p\sin{i}$ degeneracy as radial-velocity planets, allowing a more direct test of the mass--eccentricity relation. 
The sample has since grown to 85 planets (as of December 2025), with the observed correlation becoming even more pronounced.

This positive mass--eccentricity trend warrants further scrutiny, particularly to test whether its shape for warm Jupiters is consistent with excitation by planet--planet scattering. 
In this scenario, warm Jupiters are expected to form in compact, multi-planet systems, and have their eccentricities excited through dynamical interactions with nearby companions. 
If the morphology of the mass--eccentricity relation for warm Jupiters mirrors that of cold Jupiters \citep[e.g.,][]{Frelikh19}, it may point toward a unified origin for both types of gas giants. Moreover, if planet--planet scattering can excite sufficiently high eccentricities among some warm and cold Jupiters, the same dynamical process could also produce the progenitors of hot Jupiters, suggesting an overarching, unified framework for all currently-observed exo-Jupiters.  

In this work, we quantitatively characterize the mass--eccentricity relation for warm Jupiters and assess its consistency with the theoretical expectations of planet--planet scattering.
We describe the construction of the warm Jupiter sample used in this study in Section~\ref{sec:sample}. 
In Section~\ref{sec:analysis}, we present our analysis, demonstrating that the observed mass--eccentricity correlation among warm Jupiters arises naturally from planet--planet scattering. 
In Section~\ref{sec:discussion}, we explore the broader implications of this dynamical picture for the observed properties of warm Jupiter systems and the prospects for future observations.
We present our conclusions in Section~\ref{sec:conclusion}.

\section{Warm Jupiter Sample} \label{sec:sample}

\begin{figure*}
    \includegraphics{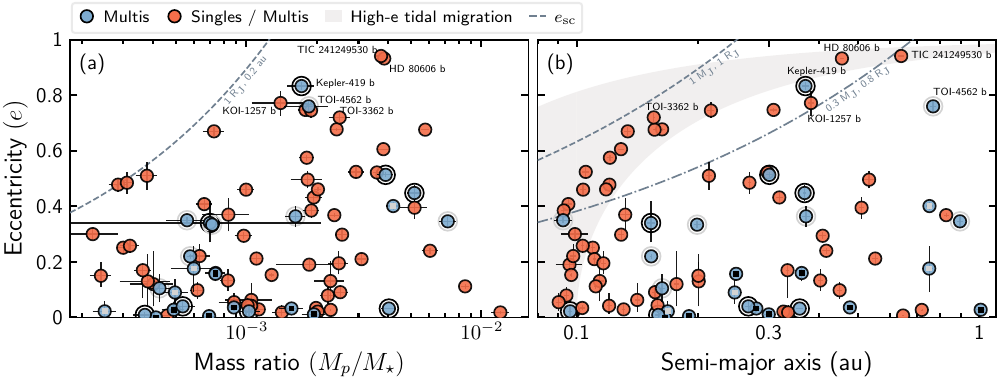}
    \caption{
    Panel (a): Warm Jupiter eccentricity as a function of planet--star mass ratio. 
    Panel (b): Warm Jupiter eccentricity as a function of semimajor axis. 
    Blue circles denote warm Jupiters with confirmed planetary companions, while orange circles indicate those in single or uncertain system architectures. 
    Additional symbols mark the type of companions: circles for Jovian planets ($M_c \geqslant 0.3\,M_{\rm Jup}$) and squares for sub-Jovian planets ($M_c < 0.3\,M_{\rm Jup}$).
    The color of these symbols indicates the companion separation, with black corresponding to $< 10$ mutual Hill radii and grey to $\geqslant 10$ mutual Hill radii.
    Mutual Hill radii are computed using companion masses inferred from transit-timing variations (TTVs) or RVs. Companion masses from RVs retain the usual $M_c\sin{i}$ degeneracy.
    Error bars show $1\sigma$ measurement uncertainties.
    The dashed and dash-dotted curves represent the maximum eccentricity expected from planet--planet scattering, assuming scattering of equal-mass planets with the additional properties marked on the curve. 
    The shaded grey region marks the parameter space consistent with high-eccentricity tidal migration, bounded by tidal dissipation efficiency and the Roche limit.
    \label{fig:obs_e}}
\end{figure*}

We construst the warm Jupiter sample from the NASA Exoplanet Archive as of December 29, 2025 \citep[][DOI: \href{https://doi.org/10.26133/NEA12}{10.26133/NEA13}]{Christiansen25}.
We select planets with reported masses, specifically true masses $M_p$ rather than $M_p\sin i$ or values inferred from mass--radius relations, as well as measured radii, eccentricities, and associated uncertainties.
We then restrict the sample to planets with masses between 0.3 and 15~$M_{\rm Jup}$ and orbital periods between 10 and 365 days.
This selection yields 85 planets in 84 systems.
The sample is dominated by discoveries from \tess and \kepler, with 51 warm Jupiters from \tess and 25 from \kepler.
We further examine the literature to verify the consistency of the reported stellar and planetary properties.
The sample is composed of FGK dwarfs with stellar masses ranging from 0.6 to 1.8\,M$_\odot$, with a median of $1.1 \pm 0.2$\,M$_\odot$.
A complete table of the selected systems, their properties, and relevant references is provided in Table~\ref{tbl:catalog}.
Figure~\ref{fig:obs_e} shows the distributions of planet mass versus eccentricity and semimajor axis versus eccentricity for the transiting warm Jupiter sample.

In terms of sample selection, we exclude systems with no or poorly-constrained eccentricities. This includes nine systems for which the reported eccentricity is fixed to zero, meaning eccentricity is not treated as a free parameter in the modeling, or for which only an upper limit is reported. This choice may reduce the number of warm Jupiters with zero eccentricity in the sample. However, since our analysis focuses on the upper eccentricity envelope, defined as the maximum eccentricity as a function of planet mass, this selection bias does not affect our interpretation.

We next consider possible biases arising from detection and follow-up observations. Because the sample consists of transiting planets, eccentric orbits have a larger geometric transit probability and are therefore more likely to be detected. Again, since our analysis focuses on the upper eccentricity envelope, this bias is beneficial, as it increases the likelihood that the sample includes planets near the edge of the envelope. In addition, transit surveys preferentially detect shorter period systems. However, Figure~\ref{fig:obs_e} shows that our warm Jupiter sample spans a broad range of semimajor axes, from approximately $0.08$ to $1\,\mathrm{au}$.
For follow-up observations, a Jupiter-mass planet at $1\,\mathrm{au}$ around a Solar-mass star induces a radial-velocity amplitude of $\sim 28\,\mathrm{m\,s^{-1}}$, which is readily detectable with modern spectrographs. Precise mass measurements are generally limited to host stars with low projected rotational velocities \citep[$v \sin i \lesssim 20\,\mathrm{km\,s^{-1}}$;][]{Bouchy01}.

We further note that the planets without known companions in our sample (orange circles in Figure~\ref{fig:obs_e}) are concentrated at smaller semi-major axes. This feature of the data may be entirely a selection effect, since many of the short-period planets in the sample are \tess detections without long-baseline observations sensitive enough to detect companions. Therefore, we emphasize that these may be either single-planet or multi-planet systems.

\begin{figure*}
    \includegraphics{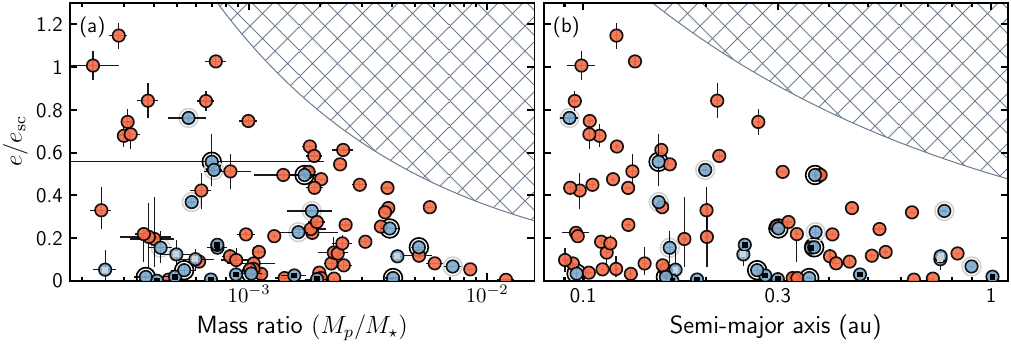}
    \caption{
    Similar to Figure~\ref{fig:obs_e}, but now showing $e / e_{\rm sc}$ on the vertical axis. With this normalization, the previously observed paucity of high-eccentricity warm Jupiters at low mass ratios or small semimajor axes is no longer apparent. The hatched regions mark the forbidden parameter space where $e > 1$, computed for a Jupiter-radius planet at $0.2\,\mathrm{au}$ in panel (a) and for a Jupiter-mass, Jupiter-radius planet in panel (b).
    \label{fig:obs_eoemax}}
\end{figure*}

\section{Analysis} \label{sec:analysis}

\subsection{Basics of Planet--Planet Scattering}

Planet--planet scattering can be understood as a relaxation process within a multi-planet system toward long-term stability.
Gravitational encounters between planets excite orbital eccentricities and inclinations, ultimately causing close encounters.
The outcomes of these close encounters depend on the Safronov number $\Theta$, with planets either undergoing collisions or being ejected from the system.
The Safronov number $\Theta$ \citep[e.g., Equation 9.46 in][]{tremaine23} is defined as:
\begin{equation}
    \Theta = \frac{v^2_{\rm esc}}{v^2_{\rm Kep}} = \left( \frac{2 G M_p}{R_p} \right) \left( \frac{a}{GM_\star} \right) = 2\,\frac{M_p}{M_\star}\frac{a}{R_p},
\end{equation}
where $v_{\rm esc}$ is the escape velocity from the surface of the planet, $v_{\rm Kep}$ is the circular Keplerian orbital velocity at the semimajor axis of the planet (comparable to the escape velocity from the star at this distance), $G$ is gravitational constant, $R_p$ is planetary radius, $M_\star$ is the stellar mass, and $a$ is the planet's semimajor axis. 
Our default assumption is that the perturber has the same mass and radius as the scattered planet when defining $v_{\rm esc}$; we relax this assumption later in our analysis. 
Systems with $\Theta < 1$ are typically collision-dominated, whereas systems with $\Theta > 1$ tend to be ejection-dominated.  

These regimes may be qualitatively understood as follows: The gravitationally-focused impact parameter $b$ required for strong scattering is approximately $b \sim GM_p/v_{\rm rel}^{2}$, where $v_{\rm rel}\sim ev_{\rm Kep}$ is the typical relative velocity of planet-planet encounters between two similar planets.  As long as $v_{\rm rel} < v_{\rm esc}$, this impact parameter exceeds the planet radius and scattering is more likely than collision.  For $\Theta > 1$, planets begin to be ejected once excited to eccentricities corresponding to $v_{\rm rel} \sim v_{\rm Kep} < v_{\rm esc}$. For $\Theta < 1$, in contrast, mutual scatterings lead to $v_{\rm rel} = v_{\rm Kep} < v_{\rm esc}$, at which point collisions commence and typical relative velocities (equivalently eccentricities) stop rising.

The maximum statistically-achievable eccentricity resulting from a population of  planets experiencing mutual scattering events is hence characterized by the relation \citep[e.g.,][]{goldreich04}:
\begin{equation}
    e_{\rm sc} = \sqrt{\Theta} = \frac{v_{\rm esc}}{v_{\rm Kep}}. \label{eqn:escattering}
\end{equation}
Limiting the eccentricity to $<1$, the maximum eccentricity for a given planet is
\begin{equation}
    e_{\rm max} = \min\left(\sqrt{\Theta}, 1\right) = \min\left(\frac{v_{\rm esc}}{v_{\rm Kep}}, 1\right). \label{eqn:emax}
\end{equation}
Over time, the chaotic evolution that occurs during an epoch of planet--planet scattering sculpts an initially unstable configuration into a dynamically more stable arrangement. Stability is achieved either through planet--planet collisions (statistically favorable once typical planetary eccentricities have been excited to $\sim e_{\rm sc}$ for $e_{\rm sc} < 1$) or through planetary ejections (statistically favorable for $e_{\rm sc} > 1$). The resulting number of planets is reduced in stable systems and resulting planets generally have eccentric and inclined orbits.

The quantity $e_{\rm sc}$ depends on both the planet--star mass ratio and the semimajor axis, placing warm Jupiters in a dynamical regime distinct from other exoplanet populations. 
Short-period small planets (rocky planets and sub-Neptunes) are typically collision-dominated due to their low masses and small semi-major axes (i.e., $\Theta < 1$), while cold Jupiters are ejection-dominated given their large masses and semi-major axes (i.e., $\Theta > 1$).
In contrast, the dynamical outcomes for warm Jupiters can be more precarious as $\Theta \sim 1$. For a representative warm Jupiter at $0.25~\mathrm{au}$ with $M_p / M_\star \sim 10^{-3}$ and $a / R_p \sim 0.25 \times 2000$, the Safronov number is $\Theta \sim 1$. 
Consequently, variations of only a factor of a few in mass or semimajor axis can shift a warm Jupiter between collisional and ejection-dominated regimes. 
Assuming a Solar-mass host star and a semimajor axis of $a = 0.25\,\mathrm{au}$, we find a scattering eccentricity of $e_{\rm sc} \simeq 0.56$ for a $0.3\,M_{\rm Jup}$ planet, increasing to $e_{\rm sc} \simeq 1$ for a $1\,M_{\rm Jup}$ or more massive planet. For a Jupiter-mass planet, the scattering threshold is $e_{\rm sc} \simeq 0.65$ at $a = 0.10\,\mathrm{au}$, corresponding to an orbital period of 11.6 days, and reaches $e_{\rm sc} \simeq 1$ by $a = 0.25\,\mathrm{au}$, or an orbital period of 45.7 days, and beyond.
This transition is imprinted in the observed eccentricity distributions as a function of both mass and semimajor axis, and, as we will argue, underlies the observed mass--eccentricity relation of warm Jupiters.

\subsection{Observation Trends}

Figure~\ref{fig:obs_e} shows warm Jupiter observations, where eccentricity is plotted against planet--star mass ratio in panel (a) and against semimajor axis in panel (b). 
There is an absence of high-eccentricity warm Jupiters at low mass ratios, shown in panel (a), similar to the finding of \citet{2024Natur.632...50G} in mass--eccentricity space.
We interpret this as a natural outcome of planet--planet scattering, which imposes a mass-dependent upper envelope on eccentricity.
To illustrate this, we compute the expected maximum eccentricity by scattering $e_{\rm sc}$ from Equation~(\ref{eqn:escattering}), for a representative planet with $a = 0.2\,\mathrm{au}$ and $R_p = 1\,R_{\rm Jup}$. 
The resulting dashed line in panel (a) traces the upper envelope of the observed distribution, especially at $M_p/M_\star < 10^{-3}$ with some deviation at higher mass ratios.
Since real systems span a range of semimajor axes and radii, this line should be viewed as a representative example rather than a strict boundary.

We now check the dependence of maximum eccentricity on semimajor axes based on the scattering theory.
Panel (b) of Figure~\ref{fig:obs_e} shows a relative lack of high-eccentricity warm Jupiters at small $a$. 
The dashed line shows $e_{\rm sc}$ for a representative Jupiter with $M_p = 1\,M_{\rm Jup}$ and $R_p = 1\,R_{\rm Jup}$ and the dash-dotted line shows for a Saturn with $M_p = 0.3\,M_{\rm Jup}$ and $R_p = 0.8\,R_{\rm Jup}$. 
While the general trend is consistent with expectations from scattering theory, the interpretation here is complicated by tidal interactions.
Planet--star tides can damp eccentricity for close-in planets, particularly those with small periastron distances. 
The shaded grey region shows the parameter space in which high-eccentricity tidal migration may operate, assuming orbital evolution along a constant angular momentum track, i.e., $a(1-e^2) = \mathrm{const}$. The upper boundary is set by the Roche limit where $a(1-e^2) = 0.034$\,au \citep{Guillochon11} and the lower boundary is set by threshold for efficient tidal circularization where we adopt $a(1-e^2) = 0.1$\,au as a representative value.
Thus, close-in warm Jupiters that were once on highly eccentric orbits may have already circularized.

We note that warm Jupiters that lie within the high-eccentricity tidal migration track are predominantly singles. They could be truly single-planet systems, or they may have companions that are so detached that they are difficult to detect. Since the majority of these systems are recent discoveries by \tess and lack long-term observing baselines, we cannot land on either scenario.
These planets may have originated at wider orbital separations and be in the process of circularizing toward tighter orbits, eventually becoming hot Jupiters, as in the classical picture of proto-hot Jupiters undergoing high-$e$ migration. 
Alternatively, within the scattering framework discussed here, it is possible that they were scattered onto highly eccentric orbits near their current locations by neighboring planets \citep{Dawson13}, with the latter either being ejected from the system or merged, such that the final observed planet appears isolated.
We discuss the potential for generating proto-hot Jupiters through planet--planet scattering in Section~\ref{sec:discussion}.

\begin{figure*}[ht]
    \centering
    \includegraphics[width=\textwidth]{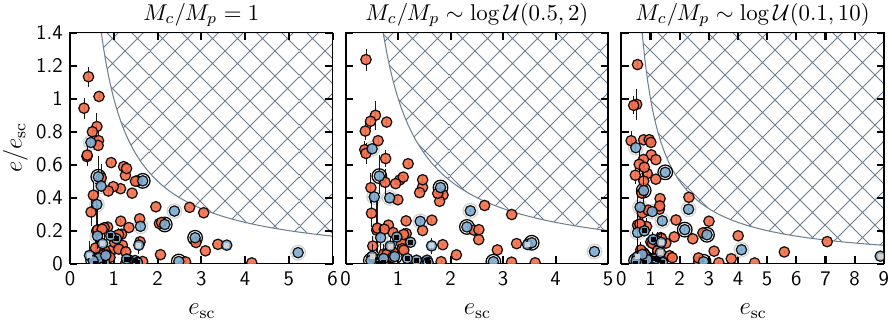}
    \caption{
    $e/e_{\rm sc}$ as a function of $e_{\rm sc}$ for different companion mass assumptions: $M_c/M_p = 1$, $M_c/M_p \sim \log U(0.5,2)$, and $M_c/M_p \sim \log U(0.1,10)$. Symbols are the same as those used in Figure~\ref{fig:obs_e}. The hatched region indicates the nonphysical regime with $e > 1$. In all cases, the distributions are nearly uniform in $e/e_{\rm sc}$, showing no clear dependence on $e_{\rm sc}$. \label{fig:escats}}
\end{figure*}

In Figure~\ref{fig:obs_eoemax}, we replot the data with the observed eccentricities normalized by $e_{\rm sc}$, the maximum eccentricity attainable through planet--planet scattering.
This normalization should remove the dependence of eccentricity on planet--star mass ratio and semimajor axis if the data is explained by the theory.
Because $e_{\rm sc}$ depends on both the scattered planet and the perturber's mass and radius, and the properties of the perturber are generally unknown, we assume for simplicity that the perturber has the same properties as the observed planet.
A more general treatment will be presented in Figure~\ref{fig:escats}.
As shown in Figure~\ref{fig:obs_eoemax}, the previously noted paucity of high-eccentricity planets at low mass ratios or small semimajor axes is no longer evident.
The hatched region marks the forbidden region where $e > 1$, calculated for a Jupiter-radius planet at $0.2\,\mathrm{au}$ in panel (a) and for a Jupiter-mass, Jupiter-radius planet in panel (b).
With this normalization, the eccentricity distribution appears approximately uniform across both mass ratio and semimajor axis, consistent with the interpretation that scattering sets the upper envelope of observed eccentricities.

The phase spaces shown in Figure \ref{fig:obs_eoemax} do not truly remove all dependencies on individual parameters that contribute to scattering physics, and the indicated forbidden zone ($e > 1$) is only illustrative as we had to choose a nominal orbital distance and a nominal mass ratio for panels a and b, respectively. 
In order to truly remove all dependencies, we plot $e/e_{\rm sc}$ vs.~$e_{\rm sc}$ in Figure \ref{fig:escats}. 
Such a plotting would also more clearly divide the systems between those that are expected to be ejection-dominated ($e_{\rm sc} > 1$) from those that are collision-dominated ($e_{\rm sc} < 1$).
The leftmost panel plots the raw data assuming equal mass perturber, and we notice two potential paucities: the lack of high $e/e_{\rm sc} \gtrsim 0.1$ planets in the nominally ejection-dominated regime ($e_{\rm sc} >$ 3); and a gap at $e/e_{\rm sc} \sim 0.3$ for $e_{\rm sc} < 2$.

While such paucities are intriguing if real, we explore the possibility that they are artifacts of our choice of perturber properties.
For a given companion mass, the mutual escape velocity is  
\begin{equation}
    v_{\rm esc} = \sqrt{\frac{2 G (M_p + M_c)}{R_p + R_c}} ,
\end{equation}
where $M_p$ and $M_c$ are the observed and companion planet masses, and $R_p$ and $R_c$ their radii. 
In addition to the equal mass perturber we explored above, we consider two other scenarios, shown in Figure~\ref{fig:escats}: 
perturber masses drawn from a log-uniform distribution spanning a factor of two smaller or larger than the observed planet, $M_c / M_p \sim \log \mathcal{U}(0.5,2)$; and perturber masses drawn from a broader distribution spanning a factor of ten smaller or larger, $M_c / M_p \sim \log \mathcal{U}(0.1,10)$. 
The corresponding radii are computed using the empirical mass--radius relation of \citet{Thorngren19}.

We find that the potential $e/e_{\rm sc}$ gap at $\sim 0.3$ largely disappears when we allow for a variation in the perturber mass.
A Saturn-mass perturber decreases $e_{\rm sc}$ owing to its smaller $M/R$, while a super-Jupiter perturber increases $e_{\rm sc}$ because of its larger $M/R$ so it is expected that some distribution in perturber masses would fill up small cavities in the data. 
The paucity of higher $e/e_{\rm sc}$ at $e_{\rm sc} > 3$ is more persistent. 
With tighter perturber mass distribution (the middle panel of Figure \ref{fig:escats}), $e_{\rm sc}$'s will have a correspondingly tighter distribution, effectively pulling in the data points at the lower and the upper edges of $e_{\rm sc}$ in the $e/e_{\rm sc}$-$e_{\rm sc}$ space. This effective concentration of data can help remove both the $e/e_{\rm sc}$ gap and the high $e/e_{\rm sc}$ paucity. With broader perturber mass distribution (the right panel of Figure \ref{fig:escats}), the distribution of $e_{\rm sc}$ will also be broadened. 
While systems originally with high $e_{\rm sc}$ would still have low $e/e_{\rm sc}$ values, systems originally with low $e_{\rm sc}$ but high $e$ can now have high $e_{\rm sc}$ if their companions are super Jupiters, and thus fill up the high $e/e_{\rm sc}$ space.
As shown in Figure \ref{fig:escats}, however, neither solution clearly removes the high $e/e_{\rm sc}$ paucity, particularly because of the small number of data points there.
With the exception of the apparent lack of high eccentricity super-Jupiters (which itself could materialize from small number statistics), we conclude that the scattering physics can generally explain the observed eccentricity--mass--orbital-distance distribution of warm Jupiters within a plausible range of perturber properties.

\section{Discussion} \label{sec:discussion}

\subsection{Scattering in the context of warm Jupiter observations}
We outline a physical picture in which planet--planet scattering explains the mass--eccentricity relation of warm Jupiters.
This framework is broadly consistent with current observational constraints, including the eccentricity bimodality, the mass--eccentricity relation discussed in this work, and the observed spin-orbit angles of warm Jupiters.

Warm Jupiters exhibit an eccentricity bimodality, with a low-$e$ component centered at $e \simeq 0.17$ and a high-$e$ component centered at $e \simeq 0.48$ \citep{Dong2021}.
Within the scattering framework, this bimodality can be understood as either (1) a transition between systems with fewer giant planets to begin with (and hence fewer planet-planet interactions) and systems with sufficiently many giant planets to reach statistical equilibrium via interactions or (2) a regime transition from collision-dominated systems to ejection-dominated systems.
The latter interpretation is further supported by trends in the joint eccentricity--mass--orbital-distance space \citep{2024Natur.632...50G}, as discussed in this work.

Warm Jupiters are also frequently observed to have low stellar obliquities \citep{2022AJ....164..104R, Wang24}.
In the scattering picture, warm Jupiters experience dynamical excitation near their current semimajor axes. At these relatively small orbital distances, inclination excitation is limited, naturally explaining why many warm Jupiters exhibit low mutual inclinations and low projected spin--orbit angles.

\subsection{Companions of warm Jupiters}
About one third of warm Jupiters in our sample have detected planetary companions, identified through direct transit detections, transit-timing variations (TTVs), or RV follow-up. 
The presence of these companions provides additional support for the scattering interpretation of warm Jupiter dynamics.
We emphasize, however, that companion searches for \tess\ warm Jupiters remain highly incomplete.
Unlike \kepler, which benefited from a nearly continuous observing baseline, the sector-based observing strategy of \tess\ makes the detection of planetary companions significantly more challenging.
Roughly speaking, \tess\ companion searches are close to complete only for planets with radii $\gtrsim 4\,R_\oplus$ and orbital periods $\lesssim 10$ days, compared to $\gtrsim 2\,R_\oplus$ and $\lesssim 50$ days for \kepler.

From \kepler, $37.2^{+18.6}_{-16.3}\%$ of warm Jupiters are found to host additional planets, the majority of which are inner super-Earths \citep{Huang16}.
Consistent with this result, Figure~\ref{fig:obs_e} shows that super-Earth companions are preferentially associated with low-eccentricity, low-mass ($M_{\rm p} < 2\,M_{\rm Jup}$) warm Jupiters.
This trend can be naturally explained by a survivor bias arising from dynamical interactions: as a warm Jupiter becomes more massive or attains a higher eccentricity, it becomes increasingly likely to eject or collide with nearby super-Earth companions.
As a result, the intrinsic companion fraction of warm Jupiters is likely higher than the fraction inferred from \kepler\ observations.

We also identify a second population of companions consisting of massive planets that may reside either close to or farther from the warm Jupiter.
Several systems host nearby Jovian companions in or near mean-motion resonance, including TOI-216 \citep{2021AJ....161..161D} and KOI-134 \citep{2025NatAs...9.1317N}.
In addition, \tess\ has revealed multiple warm Jupiter systems with pairs of giant planets exhibiting period ratios close to commensurability, such as TOI-2202 \citep{Trifonov21} and TOI-2525 \citep{2023AJ....165..179T}.
These systems suggest that convergent disk migration or related processes forming compact, resonant giant-planet configurations, providing suitable initial conditions for subsequent scattering.
If this picture is correct, an additional mechanism, such as disk dissipation or dynamical instabilities triggered by scattering, is required to move planets out of resonance.
The fraction of warm Jupiter systems that were initially captured into mean-motion resonance, and whether resonant configurations leave observable imprints on dynamically evolved systems, remain open questions.

More distant giant-planet companions are also commonly found around warm Jupiters.
This population may arise naturally from planet--planet scattering, which reduces the number of planets in a system and increases their mutual separations.
At present, however, constraints on the occurrence rate and orbital properties of such distant companions remain limited.
Ongoing and upcoming RV follow-up campaigns targeting warm Jupiters will be crucial for characterizing their companion demographics.

\subsection{Proto-hot Jupiters can gain their eccentricity from scattering}
\tess\ has revealed an emerging population of proto-hot Jupiters on high-eccentricity tidal migration tracks. 
As shown in the right panel of Figure~\ref{fig:obs_e}, more than a dozen warm Jupiters with semimajor axes between $0.1$ and $0.2\,\mathrm{au}$ lie along this track. 
These planets currently appear to be single, likely an outcome of chaotic dynamical interactions, although robust follow-up observations will be necessary to confirm their true multiplicity. 
One such example is TOI-3362\,b, a $4.0 \pm 0.4\,M_{\rm Jup}$ planet on an $18.1$-day orbit with an eccentricity of $0.720 \pm 0.016$ \citep{2021ApJ...920L..16D, 2023ApJ...958L..20E}. 
Within the scattering framework, these warm Jupiters could acquire their large eccentricities through planet--planet interactions, a picture depicted in \cite{Dawson13, Anderson20, Wu23}.

An important question is whether these planets have already undergone significant migration to reach their present semimajor axes.
Stellar obliquity provides a powerful diagnostic. 
The maximum mutual inclinations generated by scattering depend on semimajor axis, and it is generally difficult to excite mutual inclinations within the 0.1--0.2 au semimajor axis range of warm Jupiters \citep{Anderson20} despite high eccentricities. 
Thus, large stellar obliquities would suggest a migration history beyond local dynamical interactions.

Early investigations indicate that spin-orbit alignment is common among proto-hot Jupiters on the high-eccentricity migration track between 0.1--0.2 au, consistent with the planet--planet scattering pathway explored in this work. TOI-3362\,b has a highly eccentric orbit yet is spin-orbit aligned \citep{2023ApJ...958L..20E}. 
Similarly, TOI-2005\,b has an eccentric but aligned orbit \citep{Bieryla25}, as does TOI-677\,b \citep{2023AJ....166..130S, 2024AJ....167..175H, Prinoth24}. 
Furthermore, warm Jupiters within this semimajor axis range tend to exhibit spin-orbit alignment, regardless of their eccentricities \citep{2022AJ....164..104R, Wang24, 2025AJ....170...70E}.  We note that within a framework interpreting low-obliquity, high-eccentricity objects as resulting from local scattering, warm Jupiters with high eccentricities on the tidal evolution track and those with moderate eccentricities just below the track are expected to result from similar systems.  As a result, the statistical distributions of chemical compositions for these types of planets, when corrected for mass and orbital distance, are predicted to be similar.

\subsection{Initial Separation and Instability Timescale}
We have interpreted the fact that the observed population of warm Jupiters have eccentricities bounded by $e_{\rm max}$ as a signature of extended periods of planet--planet scattering (at least for those systems with eccentricities approaching the maximum). This interpretation requires each such system to have an initial population of gas giants that are closely packed for instability to occur on a timescale smaller than the age of the system.

The instability timescale \citep[e.g.,][]{chambers96, zhou07, petit20} depends on both planet mass and orbital separation between planet pairs.
Systems with large mutual Hill separations are generally stable over a star's lifetime. Therefore, the fact that most warm Jupiter systems are dynamically evolved allows us to place an upper limit on the initial separations between warm Jupiter pairs before their dynamical interactions.
\citet{zhou07} provide an empirical relation for the orbit-crossing time in the case of $M_p / M_\star = 10^{-3}$. In their simulation, they assume 9 planets with an initial semimajor separation of $\Delta$ the spacing in mutual Hill radii orbiting around a solar-mass star and assign the initial semimajor axis of the $4^{\rm th}$ planet at 1 au with an orbital period of $T_o$. The orbit crossing time $T_c$ follows:
\begin{equation}
    \log{\left(\frac{T_c}{T_o}\right)} = -5.0 + 2.2 \, \Delta,
\end{equation}
For $\Delta = 5$, this relation yields an orbit-crossing timescale of $\sim 1$ Myr.
For typical warm Jupiter systems around main-sequence stars, this scaling implies that the instability timescale does not exceed $\sim$ Gyr, corresponding to an upper limit of $\Delta \lesssim 6.6$.
Real systems, however, are more complex.
The orbit-crossing time depends on the initial eccentricity (and inclination), with higher initial $e$ leading to a shorter $T_{\rm c}$ \citep[e.g.,][]{Yoshinaga19, Pu15}. Here we adopt a simplified treatment assuming the initial $e = 0$. If the initial random velocities are in fact larger, the crossing time would be shorter, and thus the corresponding upper limit on the spacing could be larger.
Further, numerical simulations, such as those in \citet{zhou07}, assume equal-mass planets, and the effect of mass disparity on the instability timescale remains poorly understood.
In addition, the Hill radius scales with semimajor axis, so systems with the same $\Delta$ but different orbital distances may evolve differently.
Our analysis should therefore be regarded as a rough estimate of the separations between warm Jupiter pairs that can lead to dynamical instabilities consistent with observations.

\subsection{Other Considerations}
\begin{itemize}[leftmargin=*]
    \item \emph{Super-Jupiters may have lower eccentricities.} 
    Planets more massive than $5\,M_{\rm Jup}$ may appear to exhibit systematically lower eccentricities \citep{2025AJ....169....4D} or $e/e_{\rm sc}$.
    Similar evidence has been found in samples of RV-discovered planets from the California Legacy Survey \citep{blunt26} and among direct imaging planets \citep{bowler20}.
    However, the trend that these super-Jupiters have lower eccentricities is difficult to establish in our sample of warm Jupiters given (1) the small sample size and (2) the unknown role of additional, undetected perturbers, which prevents a proper characterization of $e_{\rm sc}$.   
    Accounting for these factors, we do not find significant evidence supporting a robust mass--eccentricity trend at the high-mass end.
    
    \item \emph{Concerns about tidal circularization.} 
    Eccentricity damping due to tidal dissipation can produce an eccentricity–semimajor axis correlation, with lower eccentricities at smaller semimajor axes. 
    In addition, for low-mass planets, radial-velocity surveys are more sensitive to those at short orbital periods. 
    The combination of these effects could bias detections toward low-mass, low-eccentricity planets at small semimajor axes. 
    However, most warm Jupiters in our sample do not lie on the high-eccentricity migration track, as shown in Figure~\ref{fig:obs_e}, and those that do are not preferentially low-mass.
    These considerations alleviate concerns that tidal circularization and detection bias dominate the observed mass--eccentricity trend.
    
    \item \emph{Impact of mean-motion resonance.}
    The role of mean-motion resonances in shaping warm Jupiter architectures is not addressed in this study. 
    Resonant configurations could potentially alter eccentricity and mutual inclination distributions, and quantifying what fraction of warm Jupiters participate in such resonances remains an open question. 
    Future work incorporating transit-timing variation analysis on \tess\ warm Jupiters would complement the scattering framework explored here.

    \item \emph{Stability of high-eccentricity warm Jupiters with companions.} 
    The long-term dynamical stability of eccentric warm Jupiters with detected companions has not been systematically examined.
    Systems containing planets near or within the high-eccentricity migration zone may be particularly valuable targets for detailed stability analyses and long-term $N$-body integrations.
    
\end{itemize}

\section{Conclusion} \label{sec:conclusion}

We show that planet--planet scattering naturally explains the observed mass--eccentricity relation of warm Jupiters.
In this picture, warm Jupiters originate from compact, multi-planet systems, with dynamical interactions occurring near their current semimajor axes, irrespective of the mode by which they initially arrived there, whether by convergent disk-driven migration or in-situ formation.
This framework provides a unified explanation for several observed properties of warm-Jupiter systems, including the eccentricity bimodality \citep{Dong2021}, the mass--eccentricity correlation \citep{2024Natur.632...50G}, and the low stellar obliquities of warm Jupiters at 0.1--0.3 au \citep{2022AJ....164..104R, Wang24}.
It further predicts that, beyond the known population of circular warm Jupiters with nearby companions \citep{Huang16}, eccentric warm Jupiters should frequently have additional planetary companions, many of which remain to be detected through RV and TTVs.
Moreover, for sufficiently massive systems, scattering can excite eccentricities high enough to trigger high-eccentricity migration, producing the emerging population of proto-hot Jupiters now being identified by \tess.
Taken together, these results suggest that warm Jupiters are not a distinct population, but instead represent the inner counterparts of cold Jupiters, shaped by the same underlying dynamical processes.

\begin{acknowledgments}
We thank Stephen Shectman, Songhu Wang, Yanqin Wu, and Dong Lai for insightful discussions on warm Jupiter observations that helped improve the clarity of this work.

The project was conceived at the KITP Edgeplanets program in 2025, supported in part by grant NSF PHY-2309135 to the Kavli Institute for Theoretical Physics (KITP).
EJL was supported by NSF Research Grant 2509275.

This research has made use of the NASA Exoplanet Archive, which is operated by the California Institute of Technology, under contract with the National Aeronautics and Space Administration under the Exoplanet Exploration Program.
\end{acknowledgments}

\facility{Exoplanet Archive}
% \software{}

% Catalog table
\begin{longrotatetable}
\begin{deluxetable*}{lcccccccccccl}
\tablecaption{Properties of Warm Jupiters \label{tbl:catalog}}
\tablewidth{0pt}
\tablehead{
\colhead{Planet Name} & \colhead{$M_\star$} & \colhead{$R_\star$} & \colhead{$T_{\rm eff}$} & \colhead{$M_p$} & \colhead{$R_p$} & \colhead{$P$} & \colhead{$e$} & \colhead{$\lambda$} & \colhead{TTV} & \colhead{$M_c$} & \colhead{$P_c$} & \colhead{Refs.} \\
\colhead{} & \colhead{($M_\odot$)} & \colhead{($R_\odot$)} & \colhead{(K)} & \colhead{($M_{\rm Jup}$)} & \colhead{($R_{\rm Jup}$)} & \colhead{(days)} & \colhead{} & \colhead{($^\circ$)} & \colhead{} & \colhead{($M_{\rm Jup}$)} & \colhead{(days)} & \colhead{}
}
\startdata
CoRoT-10 b & $0.89^{+0.05}_{-0.05}$ & $0.79^{+0.05}_{-0.05}$ & $5075^{+75}_{-75}$ & $2.73^{+0.14}_{-0.14}$ & $0.97^{+0.07}_{-0.07}$ & $13.241^{+0.000}_{-0.000}$ & $0.52^{+0.02}_{-0.02}$ & \nodata & N & \nodata & \nodata & 1 \\
CoRoT-9 b & $0.96^{+0.04}_{-0.04}$ & $0.96^{+0.06}_{-0.06}$ & $5625^{+80}_{-80}$ & $0.84^{+0.05}_{-0.05}$ & $1.07^{+0.07}_{-0.06}$ & $95.273^{+0.000}_{-0.000}$ & $0.13^{+0.04}_{-0.04}$ & \nodata & N & \nodata & \nodata & 2 \\
HAT-P-15 b & $1.00^{+0.23}_{-0.23}$ & $1.07^{+0.07}_{-0.07}$ & $5568^{+90}_{-90}$ & $1.94^{+0.30}_{-0.30}$ & $1.06^{+0.07}_{-0.07}$ & $10.864^{+0.000}_{-0.000}$ & $0.19^{+0.02}_{-0.02}$ & $13^{+6}_{-6}$ & N & \nodata & \nodata & 3, 4 \\
HAT-P-17 b & $0.99^{+0.15}_{-0.15}$ & $0.87^{+0.04}_{-0.04}$ & $5246^{+80}_{-80}$ & $0.58^{+0.06}_{-0.06}$ & $1.05^{+0.04}_{-0.04}$ & $10.339^{+0.000}_{-0.000}$ & $0.35^{+0.01}_{-0.01}$ & $-28^{+7}_{-7}$ & N & $3.40^{+1.10}_{-0.70}$ & $5584.0^{+7700.0}_{-2100.0}$ & 3, 4 \\
HATS-17 b & $1.13^{+0.03}_{-0.03}$ & $1.09^{+0.07}_{-0.05}$ & $5846^{+78}_{-78}$ & $1.34^{+0.07}_{-0.07}$ & $0.78^{+0.06}_{-0.06}$ & $16.255^{+0.000}_{-0.000}$ & $0.03^{+0.02}_{-0.02}$ & \nodata & N & \nodata & \nodata & 5 \\
HD 114082 b & $1.47^{+0.07}_{-0.07}$ & $1.49^{+0.05}_{-0.05}$ & $6651^{+35}_{-35}$ & $8.00^{+1.00}_{-1.00}$ & $1.00^{+0.03}_{-0.03}$ & $109.750^{+0.400}_{-0.370}$ & $0.40^{+0.04}_{-0.04}$ & \nodata & N & \nodata & \nodata & 6 \\
HD 1397 b & $1.32^{+0.04}_{-0.05}$ & $2.34^{+0.05}_{-0.06}$ & $5521^{+60}_{-60}$ & $0.41^{+0.02}_{-0.02}$ & $1.03^{+0.03}_{-0.03}$ & $11.535^{+0.001}_{-0.001}$ & $0.25^{+0.02}_{-0.02}$ & \nodata & N & \nodata & \nodata & 7 \\
HD 17156 b & $1.28^{+0.06}_{-0.06}$ & $1.52^{+0.04}_{-0.04}$ & $6046^{+76}_{-72}$ & $3.26^{+0.11}_{-0.11}$ & $1.09^{+0.03}_{-0.03}$ & $21.216^{+0.000}_{-0.000}$ & $0.68^{+0.00}_{-0.00}$ & $10^{+5}_{-5}$ & N & \nodata & \nodata & 8, 9 \\
HD 80606 b & $1.02^{+0.04}_{-0.04}$ & $1.04^{+0.03}_{-0.03}$ & $5574^{+72}_{-72}$ & $4.12^{+0.10}_{-0.10}$ & $1.00^{+0.02}_{-0.02}$ & $111.437^{+0.000}_{-0.000}$ & $0.93^{+0.00}_{-0.00}$ & $42^{+8}_{-8}$ & N & \nodata & \nodata & 1, 10 \\
K2-114 b & $0.86^{+0.04}_{-0.03}$ & $0.83^{+0.02}_{-0.02}$ & $4899^{+59}_{-58}$ & $2.01^{+0.12}_{-0.12}$ & $0.94^{+0.03}_{-0.03}$ & $11.391^{+0.000}_{-0.000}$ & $0.08^{+0.03}_{-0.03}$ & \nodata & N & \nodata & \nodata & 11 \\
K2-115 b & $0.92^{+0.04}_{-0.05}$ & $0.85^{+0.02}_{-0.02}$ & $5870^{+170}_{-150}$ & $1.01^{+0.22}_{-0.23}$ & $1.05^{+0.03}_{-0.03}$ & $20.273^{+0.000}_{-0.000}$ & $0.06^{+0.06}_{-0.04}$ & \nodata & N & \nodata & \nodata & 11 \\
K2-139 b & $0.92^{+0.03}_{-0.03}$ & $0.86^{+0.03}_{-0.03}$ & $5340^{+110}_{-110}$ & $0.39^{+0.08}_{-0.07}$ & $0.81^{+0.03}_{-0.03}$ & $28.382^{+0.000}_{-0.000}$ & $0.12^{+0.12}_{-0.08}$ & $-14^{+2}_{-3}$ & N & \nodata & \nodata & 12, 13 \\
K2-232 b & $1.19^{+0.03}_{-0.03}$ & $1.16^{+0.02}_{-0.02}$ & $6154^{+60}_{-60}$ & $0.40^{+0.04}_{-0.04}$ & $1.00^{+0.02}_{-0.02}$ & $11.168^{+0.000}_{-0.000}$ & $0.26^{+0.03}_{-0.03}$ & $-11^{+7}_{-7}$ & N & \nodata & \nodata & 14, 15 \\
K2-287 b & $1.06^{+0.02}_{-0.02}$ & $1.07^{+0.01}_{-0.01}$ & $5695^{+58}_{-58}$ & $0.32^{+0.03}_{-0.03}$ & $0.85^{+0.01}_{-0.01}$ & $14.893^{+0.000}_{-0.000}$ & $0.48^{+0.03}_{-0.03}$ & $23^{+12}_{-13}$ & N & \nodata & \nodata & 16 \\
K2-419 A b & $0.56^{+0.02}_{-0.02}$ & $0.54^{+0.02}_{-0.02}$ & $3711^{+88}_{-88}$ & $0.62^{+0.05}_{-0.05}$ & $0.94^{+0.03}_{-0.03}$ & $20.358^{+0.000}_{-0.000}$ & $0.04^{+0.04}_{-0.03}$ & \nodata & N & \nodata & \nodata & 17 \\
K2-99 b & $1.44^{+0.03}_{-0.03}$ & $2.55^{+0.02}_{-0.02}$ & $6069^{+92}_{-92}$ & $0.87^{+0.02}_{-0.02}$ & $1.06^{+0.01}_{-0.01}$ & $18.248^{+0.000}_{-0.000}$ & $0.22^{+0.01}_{-0.01}$ & \nodata & N & $8.40^{+0.20}_{-0.20}$ & $522.2^{+1.4}_{-1.4}$ & 18 \\
KOI-12 b & $1.50^{+0.10}_{-0.10}$ & $1.40^{+0.07}_{-0.06}$ & $6635^{+71}_{-71}$ & $1.10^{+3.50}_{-0.80}$ & $1.23^{+0.06}_{-0.05}$ & $17.855^{+0.000}_{-0.000}$ & $0.34^{+0.08}_{-0.07}$ & \nodata & Y & $22.00^{+7.00}_{-5.00}$ & $2500.0^{+2400.0}_{-700.0}$ & 19 \\
KOI-1257 b & $0.99^{+0.05}_{-0.05}$ & $1.13^{+0.14}_{-0.14}$ & $5520^{+80}_{-80}$ & $1.45^{+0.35}_{-0.35}$ & $0.94^{+0.12}_{-0.12}$ & $86.648^{+0.000}_{-0.000}$ & $0.77^{+0.04}_{-0.04}$ & \nodata & N & \nodata & \nodata & 20 \\
KOI-134 b & $1.41^{+0.04}_{-0.04}$ & $1.67^{+0.02}_{-0.02}$ & $6160^{+130}_{-130}$ & $1.09^{+0.12}_{-0.08}$ & $1.07^{+0.02}_{-0.02}$ & $67.128^{+0.004}_{-0.006}$ & $0.16^{+0.02}_{-0.03}$ & \nodata & Y & $0.22^{+0.02}_{-0.02}$ & $34.0^{+0.0}_{-0.0}$ & 21 \\
KOI-351 h & $1.24^{+0.10}_{-0.10}$ & $1.19^{+0.03}_{-0.03}$ & $6031^{+35}_{-35}$ & $0.64^{+0.05}_{-0.05}$ & $1.00^{+0.03}_{-0.03}$ & $331.603^{+0.000}_{-0.000}$ & $0.03^{+0.00}_{-0.00}$ & \nodata & Y & $0.05^{+0.00}_{-0.00}$ & $210.7^{+0.0}_{-0.0}$ & 22 \\
KOI-3680 b & $1.01^{+0.07}_{-0.16}$ & $0.96^{+0.05}_{-0.06}$ & $5830^{+100}_{-100}$ & $1.93^{+0.19}_{-0.21}$ & $0.99^{+0.06}_{-0.07}$ & $141.242^{+0.000}_{-0.000}$ & $0.50^{+0.03}_{-0.03}$ & \nodata & N & \nodata & \nodata & 23 \\
KOI-94 d & $1.28^{+0.05}_{-0.05}$ & $1.52^{+0.14}_{-0.14}$ & $6182^{+58}_{-58}$ & $0.33^{+0.04}_{-0.04}$ & $1.00^{+0.10}_{-0.10}$ & $22.343^{+0.000}_{-0.000}$ & $0.02^{+0.04}_{-0.04}$ & \nodata & Y & $0.05^{+0.02}_{-0.05}$ & $10.4^{+0.0}_{-0.0}$ & 24 \\
Kepler-111 c & $1.12^{+0.06}_{-0.07}$ & $1.15^{+0.04}_{-0.04}$ & $5914^{+81}_{-82}$ & $0.70^{+0.14}_{-0.13}$ & $0.63^{+0.02}_{-0.02}$ & $224.778^{+0.000}_{-0.000}$ & $0.18^{+0.08}_{-0.09}$ & \nodata & Y & $0.01$ & $3.3^{+0.0}_{-0.0}$ & 25 \\
Kepler-117 c & $1.13^{+0.13}_{-0.02}$ & $1.61^{+0.05}_{-0.05}$ & $6150^{+110}_{-110}$ & $1.84^{+0.18}_{-0.18}$ & $1.10^{+0.04}_{-0.04}$ & $50.790^{+0.000}_{-0.000}$ & $0.03^{+0.00}_{-0.00}$ & \nodata & Y & $0.09^{+0.03}_{-0.03}$ & $18.8^{+0.0}_{-0.0}$ & 26 \\
Kepler-1514 b & $1.20^{+0.07}_{-0.06}$ & $1.29^{+0.03}_{-0.03}$ & $6145^{+99}_{-80}$ & $5.28^{+0.22}_{-0.22}$ & $1.11^{+0.02}_{-0.02}$ & $217.832^{+0.000}_{-0.000}$ & $0.40^{+0.01}_{-0.01}$ & \nodata & N & $0.01$ & $10.5^{+0.0}_{-0.0}$ & 27 \\
Kepler-16 b & $0.69^{+0.00}_{-0.00}$ & $0.65^{+0.00}_{-0.00}$ & $4450^{+150}_{-150}$ & $0.33^{+0.02}_{-0.02}$ & $0.75^{+0.00}_{-0.00}$ & $228.776^{+0.020}_{-0.037}$ & $0.01^{+0.00}_{-0.00}$ & \nodata & Y & \nodata & \nodata & 28 \\
Kepler-30 c & $0.99^{+0.08}_{-0.08}$ & $0.95^{+0.12}_{-0.12}$ & $5498^{+54}_{-54}$ & $2.01^{+0.16}_{-0.16}$ & $1.10^{+0.04}_{-0.04}$ & $60.323^{+0.000}_{-0.000}$ & $0.01^{+0.00}_{-0.00}$ & \nodata & Y & $0.04^{+0.00}_{-0.00}$ & $29.3^{+0.0}_{-0.0}$ & 29 \\
Kepler-419 b & $1.39^{+0.08}_{-0.07}$ & $1.74^{+0.07}_{-0.07}$ & $6430^{+79}_{-79}$ & $2.50^{+0.30}_{-0.30}$ & $0.96^{+0.12}_{-0.12}$ & $69.755^{+0.001}_{-0.001}$ & $0.83^{+0.01}_{-0.01}$ & \nodata & Y & $7.30^{+0.40}_{-0.40}$ & $675.5^{+0.1}_{-0.1}$ & 30 \\
Kepler-432 b & $1.32^{+0.10}_{-0.07}$ & $4.06^{+0.12}_{-0.08}$ & $4995^{+78}_{-78}$ & $5.41^{+0.32}_{-0.18}$ & $1.15^{+0.04}_{-0.04}$ & $52.501^{+0.000}_{-0.000}$ & $0.51^{+0.01}_{-0.01}$ & \nodata & N & $2.43^{+0.22}_{-0.24}$ & $406.2^{+3.9}_{-2.5}$ & 31 \\
Kepler-434 b & $1.20^{+0.09}_{-0.09}$ & $1.38^{+0.13}_{-0.13}$ & $5977^{+95}_{-95}$ & $2.86^{+0.35}_{-0.35}$ & $1.13^{+0.26}_{-0.18}$ & $12.875^{+0.000}_{-0.000}$ & $0.13^{+0.07}_{-0.07}$ & \nodata & N & \nodata & \nodata & 32 \\
Kepler-553 c & $0.89^{+0.05}_{-0.04}$ & $0.90^{+0.03}_{-0.02}$ & $5191^{+76}_{-78}$ & $6.70^{+0.44}_{-0.43}$ & $1.03^{+0.03}_{-0.03}$ & $328.240^{+0.000}_{-0.000}$ & $0.35^{+0.02}_{-0.02}$ & \nodata & N & $0.00^{+0.36}_{-0.00}$ & $4.0^{+0.0}_{-0.0}$ & 25 \\
Kepler-56 c & $1.32^{+0.13}_{-0.13}$ & $4.23^{+0.15}_{-0.15}$ & $4840^{+97}_{-97}$ & $0.57^{+0.07}_{-0.06}$ & $0.87^{+0.04}_{-0.04}$ & $21.402^{+0.001}_{-0.001}$ & $0.00^{+0.01}_{-0.01}$ & \nodata & Y & $0.07^{+0.01}_{-0.01}$ & $10.5^{+0.0}_{-0.0}$ & 33 \\
Kepler-643 b & $1.15^{+0.12}_{-0.12}$ & $2.69^{+0.11}_{-0.11}$ & $4908^{+45}_{-48}$ & $1.01^{+0.20}_{-0.20}$ & $0.91^{+0.03}_{-0.02}$ & $16.339^{+0.000}_{-0.000}$ & $0.37^{+0.06}_{-0.06}$ & \nodata & N & \nodata & \nodata & 34, 35 \\
Kepler-87 b & $1.10^{+0.05}_{-0.05}$ & $1.82^{+0.04}_{-0.04}$ & $5600^{+50}_{-50}$ & $1.02^{+0.03}_{-0.03}$ & $1.20^{+0.05}_{-0.05}$ & $114.736^{+0.000}_{-0.000}$ & $0.04^{+0.01}_{-0.01}$ & \nodata & Y & $0.02^{+0.00}_{-0.00}$ & $191.2^{+0.0}_{-0.0}$ & 36 \\
NGTS-11 b & $0.86^{+0.03}_{-0.03}$ & $0.83^{+0.01}_{-0.01}$ & $5050^{+80}_{-80}$ & $0.34^{+0.09}_{-0.07}$ & $0.82^{+0.03}_{-0.03}$ & $35.455^{+0.000}_{-0.000}$ & $0.13^{+0.10}_{-0.09}$ & \nodata & N & \nodata & \nodata & 37 \\
NGTS-20 b & $1.47^{+0.09}_{-0.09}$ & $1.78^{+0.05}_{-0.05}$ & $5980^{+80}_{-80}$ & $2.98^{+0.16}_{-0.15}$ & $1.07^{+0.04}_{-0.04}$ & $54.189^{+0.000}_{-0.000}$ & $0.43^{+0.02}_{-0.02}$ & \nodata & N & \nodata & \nodata & 38 \\
NGTS-30 b & $0.94^{+0.06}_{-0.06}$ & $0.91^{+0.03}_{-0.03}$ & $5455^{+80}_{-80}$ & $0.96^{+0.06}_{-0.06}$ & $0.93^{+0.03}_{-0.03}$ & $98.298^{+0.000}_{-0.000}$ & $0.29^{+0.01}_{-0.01}$ & \nodata & N & \nodata & \nodata & 39 \\
TIC 139270665 b & $1.03^{+0.05}_{-0.06}$ & $1.02^{+0.04}_{-0.03}$ & $5844^{+84}_{-82}$ & $0.46^{+0.05}_{-0.05}$ & $0.65^{+0.02}_{-0.02}$ & $23.624^{+0.030}_{-0.031}$ & $0.10^{+0.05}_{-0.05}$ & \nodata & N & $4.89^{+0.66}_{-0.37}$ & $1010.0^{+780.0}_{-220.0}$ & 40 \\
TIC 172900988 b & $1.24^{+0.00}_{-0.00}$ & $1.38^{+0.00}_{-0.00}$ & $6050^{+100}_{-100}$ & $2.96^{+0.02}_{-0.02}$ & $1.00^{+0.04}_{-0.04}$ & $200.452^{+0.011}_{-0.011}$ & $0.03^{+0.00}_{-0.00}$ & \nodata & N & \nodata & \nodata & 41 \\
TIC 237913194 b & $1.03^{+0.06}_{-0.06}$ & $1.09^{+0.01}_{-0.01}$ & $5788^{+80}_{-80}$ & $1.94^{+0.09}_{-0.09}$ & $1.12^{+0.05}_{-0.05}$ & $15.169^{+0.000}_{-0.000}$ & $0.57^{+0.01}_{-0.01}$ & \nodata & N & \nodata & \nodata & 42 \\
TIC 241249530 b & $1.27^{+0.06}_{-0.07}$ & $1.40^{+0.03}_{-0.03}$ & $6166^{+43}_{-43}$ & $4.98^{+0.16}_{-0.18}$ & $1.19^{+0.04}_{-0.04}$ & $165.772^{+0.000}_{-0.000}$ & $0.94^{+0.00}_{-0.00}$ & $164^{+9}_{-8}$ & N & \nodata & \nodata & 43 \\
TIC 279401253 b & $1.13^{+0.02}_{-0.03}$ & $1.06^{+0.01}_{-0.01}$ & $5951^{+80}_{-80}$ & $6.14^{+0.39}_{-0.42}$ & $1.00^{+0.04}_{-0.04}$ & $76.800^{+0.060}_{-0.060}$ & $0.45^{+0.03}_{-0.03}$ & \nodata & N & $8.02^{+0.18}_{-0.18}$ & $155.3^{+0.7}_{-0.7}$ & 44 \\
TIC 393818343 b & $1.08^{+0.06}_{-0.06}$ & $1.09^{+0.02}_{-0.02}$ & $5756^{+67}_{-66}$ & $4.34^{+0.15}_{-0.15}$ & $1.09^{+0.02}_{-0.02}$ & $16.249^{+0.000}_{-0.000}$ & $0.61^{+0.00}_{-0.00}$ & \nodata & N & \nodata & \nodata & 45 \\
TIC 4672985 b & $1.01^{+0.03}_{-0.03}$ & $1.15^{+0.01}_{-0.01}$ & $5757^{+72}_{-65}$ & $12.74^{+1.01}_{-1.01}$ & $1.03^{+0.07}_{-0.07}$ & $69.048^{+0.000}_{-0.001}$ & $0.02^{+0.00}_{-0.00}$ & \nodata & N & \nodata & \nodata & 46 \\
TOI-1478 b & $0.95^{+0.06}_{-0.04}$ & $1.05^{+0.03}_{-0.03}$ & $5597^{+83}_{-82}$ & $0.88^{+0.11}_{-0.12}$ & $1.07^{+0.16}_{-0.10}$ & $10.180^{+0.000}_{-0.000}$ & $0.06^{+0.03}_{-0.03}$ & $6^{+6}_{-6}$ & N & \nodata & \nodata & 47, 48 \\
TOI-1670 c & $1.21^{+0.02}_{-0.02}$ & $1.32^{+0.02}_{-0.02}$ & $6170^{+61}_{-61}$ & $0.63^{+0.09}_{-0.08}$ & $0.99^{+0.03}_{-0.03}$ & $40.750^{+0.000}_{-0.000}$ & $0.09^{+0.05}_{-0.04}$ & $-0^{+2}_{-2}$ & N & $0.13$ & $11.0^{+0.0}_{-0.0}$ & 49, 50 \\
TOI-1898 b & $1.25^{+0.03}_{-0.04}$ & $1.61^{+0.03}_{-0.03}$ & $6241^{+88}_{-85}$ & $0.41^{+0.02}_{-0.03}$ & $0.84^{+0.02}_{-0.03}$ & $45.522^{+0.000}_{-0.000}$ & $0.48^{+0.04}_{-0.04}$ & \nodata & N & \nodata & \nodata & 51, 52 \\
TOI-1899 b & $0.63^{+0.03}_{-0.03}$ & $0.61^{+0.02}_{-0.02}$ & $3926^{+45}_{-47}$ & $0.67^{+0.04}_{-0.04}$ & $0.99^{+0.03}_{-0.03}$ & $29.090^{+0.000}_{-0.000}$ & $0.04^{+0.03}_{-0.03}$ & \nodata & N & \nodata & \nodata & 53, 54 \\
TOI-2010 b & $1.11^{+0.05}_{-0.06}$ & $1.08^{+0.03}_{-0.03}$ & $5929^{+74}_{-74}$ & $1.29^{+0.06}_{-0.06}$ & $1.05^{+0.03}_{-0.03}$ & $141.834^{+0.000}_{-0.000}$ & $0.21^{+0.02}_{-0.02}$ & \nodata & N & \nodata & \nodata & 55 \\
TOI-2145 b & $1.72^{+0.03}_{-0.04}$ & $2.75^{+0.06}_{-0.05}$ & $6200^{+84}_{-81}$ & $5.56^{+0.22}_{-0.23}$ & $1.11^{+0.03}_{-0.02}$ & $10.261^{+0.000}_{-0.000}$ & $0.21^{+0.02}_{-0.02}$ & $7^{+3}_{-3}$ & N & \nodata & \nodata & 51, 56 \\
TOI-216 b & $0.77^{+0.03}_{-0.03}$ & $0.75^{+0.01}_{-0.01}$ & $5026^{+125}_{-125}$ & $0.56^{+0.02}_{-0.02}$ & $0.90^{+0.02}_{-0.02}$ & $34.526$ & $0.00^{+0.00}_{-0.00}$ & \nodata & Y & $0.06^{+0.00}_{-0.00}$ & $17.2$ & 57, 58 \\
TOI-2180 b & $1.11^{+0.05}_{-0.05}$ & $1.64^{+0.03}_{-0.03}$ & $5695^{+58}_{-60}$ & $2.75^{+0.09}_{-0.08}$ & $1.01^{+0.02}_{-0.02}$ & $260.790^{+0.590}_{-0.580}$ & $0.37^{+0.01}_{-0.01}$ & \nodata & Y & \nodata & \nodata & 59 \\
TOI-2202 b & $0.84^{+0.03}_{-0.03}$ & $0.81^{+0.02}_{-0.02}$ & $5169^{+80}_{-78}$ & $0.90^{+0.09}_{-0.10}$ & $0.98^{+0.02}_{-0.02}$ & $11.913^{+0.000}_{-0.000}$ & $0.02^{+0.02}_{-0.01}$ & $26^{+12}_{-15}$ & Y & $0.37^{+0.10}_{-0.08}$ & $24.7^{+0.0}_{-0.0}$ & 60 \\
TOI-2295 b & $1.17^{+0.07}_{-0.07}$ & $1.46^{+0.06}_{-0.06}$ & $5730^{+140}_{-140}$ & $0.88^{+0.04}_{-0.04}$ & $1.47^{+0.85}_{-0.53}$ & $30.033^{+0.000}_{-0.000}$ & $0.33^{+0.01}_{-0.01}$ & \nodata & N & $5.61^{+0.23}_{-0.24}$ & $966.5^{+4.3}_{-4.2}$ & 61 \\
TOI-2338 b & $0.99^{+0.03}_{-0.02}$ & $1.05^{+0.01}_{-0.01}$ & $5581^{+60}_{-60}$ & $5.98^{+0.21}_{-0.20}$ & $1.00^{+0.02}_{-0.02}$ & $22.654^{+0.000}_{-0.000}$ & $0.68^{+0.00}_{-0.00}$ & \nodata & N & \nodata & \nodata & 62 \\
TOI-2373 b & $1.04^{+0.03}_{-0.03}$ & $1.10^{+0.02}_{-0.02}$ & $5651^{+80}_{-80}$ & $9.30^{+0.20}_{-0.20}$ & $0.93^{+0.02}_{-0.02}$ & $13.337^{+0.000}_{-0.000}$ & $0.11^{+0.01}_{-0.01}$ & \nodata & N & \nodata & \nodata & 63 \\
TOI-2447 b & $1.03^{+0.03}_{-0.03}$ & $1.01^{+0.01}_{-0.01}$ & $5730^{+80}_{-80}$ & $0.39^{+0.03}_{-0.03}$ & $0.86^{+0.01}_{-0.01}$ & $69.337^{+0.000}_{-0.000}$ & $0.17^{+0.02}_{-0.10}$ & \nodata & Y & \nodata & \nodata & 64 \\
TOI-2449 b & $1.08^{+0.04}_{-0.05}$ & $1.06^{+0.01}_{-0.01}$ & $6021^{+62}_{-62}$ & $0.70^{+0.04}_{-0.04}$ & $1.00^{+0.01}_{-0.01}$ & $106.145^{+0.000}_{-0.000}$ & $0.10^{+0.03}_{-0.03}$ & \nodata & N & \nodata & \nodata & 65 \\
TOI-2485 b & $1.16^{+0.05}_{-0.05}$ & $1.72^{+0.07}_{-0.07}$ & $6016^{+119}_{-119}$ & $2.41^{+0.09}_{-0.09}$ & $1.08^{+0.04}_{-0.04}$ & $11.235^{+0.000}_{-0.000}$ & $0.03^{+0.01}_{-0.01}$ & $-20^{+26}_{-21}$ & N & \nodata & \nodata & 66 \\
TOI-2497 b & $1.86^{+0.09}_{-0.08}$ & $2.36^{+0.12}_{-0.11}$ & $7360^{+320}_{-300}$ & $4.82^{+0.41}_{-0.41}$ & $0.99^{+0.06}_{-0.05}$ & $10.656^{+0.000}_{-0.000}$ & $0.20^{+0.04}_{-0.04}$ & \nodata & N & \nodata & \nodata & 67 \\
TOI-2525 c & $0.85^{+0.04}_{-0.04}$ & $0.79^{+0.03}_{-0.03}$ & $5096^{+102}_{-102}$ & $0.66^{+0.03}_{-0.03}$ & $0.90^{+0.01}_{-0.01}$ & $49.252^{+0.000}_{-0.000}$ & $0.16^{+0.01}_{-0.01}$ & \nodata & Y & $0.08^{+0.01}_{-0.01}$ & $23.3^{+0.0}_{-0.0}$ & 68 \\
TOI-2529 b & $1.11^{+0.01}_{-0.02}$ & $1.70^{+0.02}_{-0.03}$ & $5802^{+60}_{-52}$ & $2.34^{+0.20}_{-0.20}$ & $1.03^{+0.05}_{-0.05}$ & $64.595^{+0.000}_{-0.000}$ & $0.02^{+0.02}_{-0.01}$ & \nodata & N & \nodata & \nodata & 46 \\
TOI-2537 b & $0.77^{+0.05}_{-0.05}$ & $0.77^{+0.04}_{-0.04}$ & $4870^{+150}_{-150}$ & $1.31^{+0.09}_{-0.09}$ & $1.00^{+0.06}_{-0.06}$ & $94.102^{+0.001}_{-0.001}$ & $0.36^{+0.04}_{-0.04}$ & \nodata & Y & $7.23^{+0.52}_{-0.45}$ & $1920.0^{+230.0}_{-140.0}$ & 61 \\
TOI-2589 b & $0.93^{+0.03}_{-0.02}$ & $1.07^{+0.01}_{-0.01}$ & $5579^{+70}_{-70}$ & $3.50^{+0.10}_{-0.10}$ & $1.08^{+0.03}_{-0.03}$ & $61.628^{+0.000}_{-0.000}$ & $0.52^{+0.01}_{-0.01}$ & \nodata & N & \nodata & \nodata & 62 \\
TOI-3362 b & $1.53^{+0.02}_{-0.02}$ & $1.75^{+0.02}_{-0.02}$ & $6800^{+100}_{-100}$ & $4.00^{+0.40}_{-0.40}$ & $1.20^{+0.20}_{-0.20}$ & $18.095^{+0.000}_{-0.000}$ & $0.72^{+0.02}_{-0.02}$ & $1^{+3}_{-3}$ & N & \nodata & \nodata & 69, 70 \\
TOI-3837 b & $0.89^{+0.05}_{-0.05}$ & $1.20^{+0.06}_{-0.05}$ & $5905^{+157}_{-155}$ & $0.59^{+0.05}_{-0.05}$ & $0.97^{+0.05}_{-0.06}$ & $11.889^{+0.000}_{-0.000}$ & $0.22^{+0.04}_{-0.05}$ & \nodata & N & \nodata & \nodata & 71 \\
TOI-4127 b & $1.23^{+0.07}_{-0.07}$ & $1.29^{+0.05}_{-0.05}$ & $6096^{+115}_{-115}$ & $2.30^{+0.11}_{-0.11}$ & $1.10^{+0.04}_{-0.03}$ & $56.399^{+0.000}_{-0.000}$ & $0.75^{+0.01}_{-0.01}$ & $4^{+17}_{-16}$ & N & \nodata & \nodata & 72, 73 \\
TOI-4406 b & $1.19^{+0.03}_{-0.03}$ & $1.29^{+0.01}_{-0.01}$ & $6219^{+70}_{-70}$ & $0.30^{+0.03}_{-0.03}$ & $1.00^{+0.02}_{-0.02}$ & $30.084^{+0.000}_{-0.000}$ & $0.15^{+0.05}_{-0.04}$ & \nodata & N & \nodata & \nodata & 62 \\
TOI-4465 b & $0.93^{+0.06}_{-0.06}$ & $1.01^{+0.04}_{-0.04}$ & $5545^{+19}_{-19}$ & $5.89^{+0.26}_{-0.26}$ & $1.25^{+0.08}_{-0.07}$ & $101.941^{+0.000}_{-0.000}$ & $0.24^{+0.01}_{-0.01}$ & \nodata & N & \nodata & \nodata & 74 \\
TOI-4504 c & $0.89^{+0.06}_{-0.04}$ & $0.92^{+0.04}_{-0.04}$ & $5315^{+60}_{-60}$ & $3.77^{+0.18}_{-0.18}$ & $0.99^{+0.05}_{-0.05}$ & $82.972^{+0.000}_{-0.000}$ & $0.03^{+0.00}_{-0.00}$ & \nodata & Y & $1.42^{+0.07}_{-0.06}$ & $40.6^{+0.0}_{-0.0}$ & 75 \\
TOI-4515 b & $0.95^{+0.02}_{-0.02}$ & $0.86^{+0.03}_{-0.03}$ & $5433^{+70}_{-70}$ & $2.00^{+0.05}_{-0.05}$ & $1.09^{+0.04}_{-0.04}$ & $15.266^{+0.000}_{-0.000}$ & $0.46^{+0.01}_{-0.01}$ & $3^{+2}_{-2}$ & N & \nodata & \nodata & 76, 13 \\
TOI-4562 b & $1.19^{+0.06}_{-0.06}$ & $1.15^{+0.05}_{-0.05}$ & $6096^{+32}_{-32}$ & $2.30^{+0.48}_{-0.47}$ & $1.12^{+0.01}_{-0.01}$ & $225.106^{+0.000}_{-0.000}$ & $0.76^{+0.02}_{-0.02}$ & \nodata & Y & $5.77^{+0.37}_{-0.56}$ & $3990.0^{+201.0}_{-192.0}$ & 77, 78 \\
TOI-4582 b & $1.34^{+0.06}_{-0.06}$ & $2.50^{+0.10}_{-0.10}$ & $5190^{+100}_{-100}$ & $0.53^{+0.05}_{-0.05}$ & $0.94^{+0.09}_{-0.12}$ & $31.034^{+0.001}_{-0.001}$ & $0.51^{+0.05}_{-0.05}$ & \nodata & N & \nodata & \nodata & 79 \\
TOI-481 b & $1.14^{+0.02}_{-0.01}$ & $1.66^{+0.02}_{-0.02}$ & $5735^{+72}_{-72}$ & $1.53^{+0.03}_{-0.03}$ & $0.99^{+0.01}_{-0.01}$ & $10.331^{+0.000}_{-0.000}$ & $0.15^{+0.01}_{-0.01}$ & \nodata & N & \nodata & \nodata & 80 \\
TOI-4914 b & $1.03^{+0.06}_{-0.06}$ & $1.00^{+0.02}_{-0.02}$ & $5805^{+62}_{-62}$ & $0.71^{+0.04}_{-0.04}$ & $1.15^{+0.03}_{-0.03}$ & $10.601^{+0.000}_{-0.000}$ & $0.41^{+0.02}_{-0.02}$ & \nodata & N & \nodata & \nodata & 81 \\
TOI-5027 b & $1.02^{+0.05}_{-0.05}$ & $0.92^{+0.04}_{-0.04}$ & $5909^{+145}_{-144}$ & $2.02^{+0.13}_{-0.13}$ & $0.96^{+0.05}_{-0.06}$ & $10.244^{+0.000}_{-0.000}$ & $0.39^{+0.03}_{-0.03}$ & $6^{+8}_{-6}$ & N & \nodata & \nodata & 71, 13 \\
TOI-5110 b & $1.47^{+0.09}_{-0.09}$ & $2.33^{+0.10}_{-0.10}$ & $6160^{+150}_{-150}$ & $2.90^{+0.13}_{-0.13}$ & $1.07^{+0.05}_{-0.05}$ & $30.159^{+0.000}_{-0.000}$ & $0.74^{+0.03}_{-0.03}$ & \nodata & N & \nodata & \nodata & 61 \\
TOI-5153 b & $1.24^{+0.07}_{-0.07}$ & $1.40^{+0.04}_{-0.04}$ & $6300^{+80}_{-80}$ & $3.26^{+0.18}_{-0.17}$ & $1.06^{+0.04}_{-0.04}$ & $20.330^{+0.000}_{-0.000}$ & $0.09^{+0.02}_{-0.03}$ & \nodata & N & \nodata & \nodata & 38 \\
TOI-5542 b & $0.89^{+0.06}_{-0.03}$ & $1.06^{+0.04}_{-0.04}$ & $5700^{+80}_{-80}$ & $1.32^{+0.10}_{-0.10}$ & $1.01^{+0.04}_{-0.04}$ & $75.124^{+0.000}_{-0.000}$ & $0.02^{+0.03}_{-0.01}$ & \nodata & N & \nodata & \nodata & 82 \\
TOI-558 b & $1.35^{+0.06}_{-0.07}$ & $1.50^{+0.04}_{-0.04}$ & $6466^{+95}_{-93}$ & $3.61^{+0.15}_{-0.15}$ & $1.09^{+0.04}_{-0.04}$ & $14.574^{+0.000}_{-0.000}$ & $0.30^{+0.02}_{-0.02}$ & $12^{+8}_{-5}$ & N & \nodata & \nodata & 83, 13 \\
TOI-6628 b & $0.97^{+0.06}_{-0.06}$ & $1.01^{+0.06}_{-0.05}$ & $5463^{+143}_{-142}$ & $0.74^{+0.06}_{-0.06}$ & $0.98^{+0.06}_{-0.06}$ & $18.184^{+0.000}_{-0.000}$ & $0.67^{+0.01}_{-0.02}$ & \nodata & N & \nodata & \nodata & 71 \\
TOI-677 b & $1.18^{+0.06}_{-0.06}$ & $1.28^{+0.03}_{-0.03}$ & $6295^{+77}_{-77}$ & $1.23^{+0.08}_{-0.08}$ & $1.17^{+0.03}_{-0.03}$ & $11.237^{+0.000}_{-0.000}$ & $0.46^{+0.02}_{-0.02}$ & $0^{+1}_{-1}$ & N & \nodata & \nodata & 84, 85, 86 \\
TOI-7510 c & $1.06^{+0.07}_{-0.07}$ & $1.03^{+0.03}_{-0.03}$ & $5720^{+50}_{-50}$ & $0.41^{+0.04}_{-0.04}$ & $0.96^{+0.03}_{-0.03}$ & $22.569$ & $0.01^{+0.01}_{-0.01}$ & \nodata & Y & $0.60^{+0.06}_{-0.06}$ & $48.9$ & 87 \\
TOI-7510 d & $1.06^{+0.07}_{-0.07}$ & $1.03^{+0.03}_{-0.03}$ & $5720^{+50}_{-50}$ & $0.60^{+0.06}_{-0.06}$ & $0.94^{+0.03}_{-0.03}$ & $48.864$ & $0.04^{+0.02}_{-0.02}$ & \nodata & Y & $0.41^{+0.04}_{-0.04}$ & $22.6$ & 87 \\
WASP-117 b & $1.29^{+0.30}_{-0.30}$ & $1.22^{+0.06}_{-0.06}$ & $6040^{+90}_{-90}$ & $0.30^{+0.05}_{-0.05}$ & $1.06^{+0.07}_{-0.07}$ & $10.022^{+0.001}_{-0.001}$ & $0.30^{+0.02}_{-0.02}$ & $-47^{+6}_{-5}$ & N & \nodata & \nodata & 3, 88 \\
\enddata
\tablecomments{Warm Jupiters with orbital periods between 10--365 days, masses between 0.3--15~$M_{\rm Jup}$, and well constrained eccentricities, masses, and radii. $M_\star$, $R_\star$, and $T_{\rm eff}$ denote the host star mass, radius, and effective temperature, respectively. $M_p$, $R_p$, $P$, $e$, and $\lambda$ denote the warm Jupiter mass, radius, orbital period, eccentricity, and projected spin orbit angle. TTV indicates whether the warm Jupiter exhibits detected transit-timing variations. $M_c$ and $P_c$ are the companion mass and orbital period when a companion is detected. If multiple companions are present, we report the one with the strongest dynamical impact, as determined by the Hill factor. References: (1) \citet{2017AA...602A.107B}, (2) \citet{2017AA...603A..43B}, (3) \citet{2017AJ....153..136S}, (4) \citet{2022AA...664A.162M}, (5) \citet{2016AJ....151...89B}, (6) \citet{2022AA...667L..14Z}, (7) \citet{2019AA...623A.100N}, (8) \citet{2023AJ....165..252K}, (9) \citet{2009PASJ...61..991N}, (10) \citet{2010AA...516A..95H}, (11) \citet{2023AJ....165..155T}, (12) \citet{2018MNRAS.475.1765B}, (13) \citet{2025AJ....170...70E}, (14) \citet{2018MNRAS.477.2572B}, (15) \citet{2021AJ....162...50W}, (16) \citet{2024AA...690A.379K}, (17) \citet{2024AJ....168..235K}, (18) \citet{2022MNRAS.510.5035S}, (19) \citet{2017AJ....154...64M}, (20) \citet{2014AA...571A..37S}, (21) \citet{2025NatAs...9.1317N}, (22) \citet{2025AJ....170..146S}, (23) \citet{2019AA...623A.104H}, (24) \citet{2013ApJ...768...14W}, (25) \citet{2024ApJS..271...16D}, (26) \citet{2015AA...573A.124B}, (27) \citet{2021AJ....161..103D}, (28) \citet{2011Sci...333.1602D}, (29) \citet{2012Natur.487..449S}, (30) \citet{2014ApJ...791...89D}, (31) \citet{2015ApJ...803...49Q}, (32) \citet{2015AA...575A..71A}, (33) \citet{2013Sci...342..331H}, (34) \citet{2016ApJ...822...86M}, (35) \citet{2018ApJ...861L...5G}, (36) \citet{2014AA...561A.103O}, (37) \citet{2020ApJ...898L..11G}, (38) \citet{2022AA...666A..46U}, (39) \citet{2024AA...686A.230B}, (40) \citet{2024AJ....167..170P}, (41) \citet{2021AJ....162..234K}, (42) \citet{2020AJ....160..275S}, (43) \citet{2024Natur.632...50G}, (44) \citet{2023ApJ...946L..36B}, (45) \citet{2024AJ....168...26S}, (46) \citet{2024AA...683A.192J}, (47) \citet{2021AJ....161..194R}, (48) \citet{2022AJ....164..104R}, (49) \citet{2022AJ....163..225T}, (50) \citet{2023ApJ...959L...5L}, (51) \citet{2023AJ....166...33M}, (52) \citet{2024ApJS..272...32P}, (53) \citet{2023AJ....166...90L}, (54) \citet{2020AJ....160..147C}, (55) \citet{2023AJ....166..239M}, (56) \citet{2025AJ....169....4D}, (57) \citet{2021AJ....161..161D}, (58) \citet{2019MNRAS.486.4980K}, (59) \citet{2022AJ....163...61D}, (60) \citet{2023AJ....166..266R}, (61) \citet{2025AA...694A..36H}, (62) \citet{2023AJ....165..227B}, (63) \citet{2023AJ....166..271E}, (64) \citet{2024MNRAS.533..109G}, (65) \citet{2025AA...703A.258U}, (66) \citet{2024AA...690A..18C}, (67) \citet{2023MNRAS.521.2765R}, (68) \citet{2023AJ....165..179T}, (69) \citet{2021ApJ...920L..16D}, (70) \citet{2023ApJ...958L..20E}, (71) \citet{2025AA...694A.268T}, (72) \citet{2023AJ....165..234G}, (73) \citet{2025AA...704A.194M}, (74) \citet{2025AJ....170...41E}, (75) \citet{2025ApJ...978L..22V}, (76) \citet{2024AA...682A.135C}, (77) \citet{2023AJ....165..121H}, (78) \citet{2024AA...690L...7F}, (79) \citet{2023AJ....165...44G}, (80) \citet{2020AJ....160..235B}, (81) \citet{2024AA...691A..67M}, (82) \citet{2022AA...668A..29G}, (83) \citet{2022AJ....163....9I}, (84) \citet{2020AJ....159..145J}, (85) \citet{2023AJ....166..130S}, (86) \citet{2024AJ....167..175H}, (87) \citet{2025AA...704L..16A}, (88) \citet{2021AA...646A.168C}.}
\end{deluxetable*}
\end{longrotatetable}

% \section{Using Chinese, Japanese, and Korean characters}
% Authors have the option to include names in Chinese, Japanese, or Korean (CJK)
% characters in addition to the English name. The names will be displayed
% in parentheses after the English name. The way to do this in AASTeX is to
% use the CJK package available at \url{https://ctan.org/pkg/cjk?lang=en}.
% Further details on how to implement this and solutions for common problems,
% please go to \url{https://journals.aas.org/nonroman/}.

\bibliography{dynamics}{}
\bibliographystyle{aasjournal}

\end{document}